\documentstyle[prl,aps,epsbox]{revtex}
\begin{document}
\draft
\preprint{}
\twocolumn[\hsize\textwidth\columnwidth\hsize\csname@twocolumnfalse%
\endcsname
\title{
Green Function Monte Carlo Method 
for Excited States of Quantum System
}
\author{
Taksu Cheon
}
\address{
Department of Physics, Hosei University, Fujimi, 
Chiyoda-ku, Tokyo 102, Japan
}
\date{September 17, 1996}
\maketitle
\begin{abstract}

A novel scheme to solve the quantum eigenvalue problem through the
imaginary-time Green function Monte Carlo method is presented.  This
method is applicable to the excited states as well as to the ground
state of a generic system.  We demonstrate the validity of the method
with the numerical examples on three simple systems including a
discretized sine-Gordon model.
\\
({\it to be published in} 
Progress of Theoretical Physics {\bf 96} vol.5 (1996))
\end{abstract}
%
%
]

The stochastic quantization has been successfully applied in recent
years to the various problems of physics ranging from quantum
chromo-dynamics to the plasmon oscillation \cite{BI92,CR79,PC94}.
While it can handle very complex system which is usually beyond the
approach either by the direct diagonalization or by the perturbation,
it has been only applicable to the calculation of the ground state
properties of the system.  In this paper, we propose a new method
based on the imaginary-time Green function Monte Carlo approach, which
is extended to the excited states of Hamiltonian systems.  We envision
the application of the method both in non-relativistic quantum
few-body problems \cite{ZK81,LP81,SW93}, and in lattice field
theories \cite{LA94}. The method is illustrated through the examples
of a Morse oscillator, of a simple schematic deformed shell model with
two interacting particles in two-dimension, and of sine-Gordon coupled
oscillators.

	Let us suppose that we want to solve an eigenstate
problem of a Hamiltonian $H$, namely
\begin{eqnarray}
\label{1}
H \left| {\psi _\alpha } \right\rangle
= E_\alpha \left| {\psi _\alpha } \right\rangle .
\end{eqnarray}
We assume that each eigenstate is normalized to unity.  The evolution
operator with imaginary time $\tau$ acts on an arbitrary state
$\phi_1$ as
\begin{eqnarray}
\label{2}
e^{-\tau  H}\left| {\phi _1} \right\rangle
=\sum\limits_\alpha  {e^{-\tau  E_\alpha }
\left| {\psi _\alpha } \right\rangle
\left\langle {\psi_\alpha } \right.\left| {\phi_1} \right\rangle} .
\end{eqnarray}
If we discretize the imaginary time $\tau$ with the unit $\Delta\tau$,
and call the state at the $n$-th step of the evolution 
$\left| {\phi _1} \right\rangle^{(n)}
\equiv \exp{-n \Delta \tau H}\left| {\phi_1} \right\rangle$, 
the evolution at each step $\Delta\tau$ is described by
\begin{eqnarray}
\label{3}
\left| {\phi _1} \right\rangle^{(n+1)}
=e^{-\Delta \tau  H} \left| {\phi_1} \right\rangle^{(n)} .
\end{eqnarray}
After sufficient number of steps, the state with the lowest
energy is filtered out: 
\begin{eqnarray} 
\label{4} 
\left|{\phi _1} \right\rangle ^{(n)} 
\to \ e^{-n \Delta \tau E_1}\left| {\psi _1} \right\rangle 
    \left\langle {\psi _1} \right|\left. {\phi _1}\right\rangle 
\ \ \ \ (n\ \to \infty) . 
\end{eqnarray}
We consider the second sequence
of states $\phi_2^{(n)}$ which satisfy the evolution equation
\begin{eqnarray} 
\label{5} 
\left| {\phi _2} \right\rangle ^{(n+1)} 
&=& e^{-\Delta \tau  H} \left| {\phi _2} \right\rangle ^{(n)}
\\ \nonumber
& & 
  - \left| {\phi _1} \right\rangle ^{(n+1)}
   {{{}^{(n)}\!\left\langle {\phi _1}\right|
   e^{- \Delta \tau  H} \left| {\phi _2} \right\rangle ^{(n)}}
    \over {{}^{(n)}\!\left\langle {\phi_1} \right|\left. {\phi_1}
    \right\rangle ^{(n+1)} }} .
\end{eqnarray}
The states $\phi_1^{(n)}$  and  $\phi_2^{(n)}$ at adjacent time steps 
satisfy the 
orthogonality 
${{}^{(n)}\!\left\langle {\phi _1} \right| 
\left. {\phi _2} \right\rangle^{(n+1)}=0}$.  
Expanding eq. (\ref{5}) in the first order of $\Delta \tau$, we have
\begin{eqnarray}
\label{6}
\left| {\phi _2} \right\rangle^{(n+1)}
&\approx&  e^{-\Delta \tau  H} \times
\\ \nonumber
& &  
\left\{ {1-{{\left| {\phi_1}
  \right\rangle^{(n)}{}^{(n)}\!\left\langle {\phi_1} \right|} 
  \over {{}^{(n)}\!\left\langle {\phi _1} \right| \left. {\phi _1} 
  \right\rangle^{(n)} }}\left[ {1+\Delta \tau (E_1-H)} \right]} 
  \right\} \left| {\phi_2} \right\rangle^{(n)} 
\\ \nonumber
&\approx& \left( {\ 1-\left| {\psi _1} \right\rangle
  \left\langle {\psi_1} \right| } \right) 
  e^{-\Delta \tau  H} \left| {\phi _2} \right\rangle^{(n)} \\ \nonumber
&=& e^{-\Delta \tau  H} \left( 
              { 1-\left| {\psi _1} \right\rangle
                  \left\langle {\psi _1} \right|} \right)
    \left| {\phi _2} \right\rangle^{(n)}
\end{eqnarray}
whose $\Delta\tau$-dependent term in the first line vanishes at the
large time step because of eq. (\ref{4}).  
It is now clear that the repeated operation of eq. (\ref{5}) 
filters out the first excited state, namely:
\begin{eqnarray}
\label{7}
\left| {\phi _2} \right\rangle^{(n)}
&\approx& 
    \left[ { \left( { 1-\left| {\psi _1} \right\rangle
                    \left\langle {\psi _1} \right| }
         \right) e^{-\Delta \tau H} }\right]^{n} 
    \left| {\phi _2} \right\rangle \\ \nonumber
&=& \left( 
         { 1-\left| {\psi _1} 
           \right\rangle\left\langle {\psi_1} \right| } 
    \right)\ e^{-n\Delta \tau H} 
    \left| {\phi _2} \right\rangle \\ \nonumber
&=& e^{-n\Delta \tau H}\ 
    \left( {\ 1-\left| {\psi_1} \right\rangle
                \left\langle {\psi _1} \right| }
    \right)
    \left| {\phi _2} \right\rangle \\ \nonumber
&\to& e^{-n \Delta \tau  E_2} 
    \left| {\psi _2} \right\rangle 
    \left\langle {\psi _2} \right| 
    \left. {\phi _2} \right\rangle  \ \ (n \to  \infty ) .
\end{eqnarray}
It is important to notice that the orthogonalization at
each time step, eq. (\ref{5}) is essential in actual calculation,
because the single Schmidt subtraction either at the starting or at
the end of the repeated iteration 
is impractical with finite accuracy numerics.  One can
generalize eqs. (\ref{3}) and (\ref{5}) to obtain the evolution
equation for the $\alpha$-th energy eigenstate 
($\alpha = 1, 2, 3, ...$) as
\begin{eqnarray}
\label{8}
\left| {\phi _\alpha } \right\rangle^{(n+1)}\ 
&=& e^{-\Delta \tau  H} \left| {\phi _\alpha } 
\right\rangle^{(n)}
\\ \nonumber
& &- \sum\limits_{\beta =1}^{\alpha -1} 
{\ \left| {\phi _\beta } 
\right\rangle^{(n+1)}{{{}^{(n)}\!\left\langle {\phi_\beta } \right|
 e^{-\Delta \tau  H} 
\left| {\phi_\alpha } \right\rangle^{(n)}} \over {{}^{(n)}\!\left\langle 
{\phi_\beta } 
\right| \left. {\phi_\beta } \right\rangle^{(n+1)}\ }}} .
\end{eqnarray}
Then, with the step-wise asymptotic orthogonal conditions
${{}^{(n)}\left\langle {\phi _\beta } \right| \left. {\phi _\alpha }
\right\rangle^{(n+1)}} \to 0$   $(n \to \infty)$ for different states
$\alpha$ and $\beta$ , one can recursively filter out the $\alpha$-th
energy eigenstate as
\begin{eqnarray}
\label{9}
\left| {\phi _\alpha } \right\rangle^{(n)}
\to  e^{-n\Delta \tau E_\alpha } 
    \left| {\psi_\alpha } \right\rangle
    \left\langle {\psi_\alpha } \right| 
    \left. {\phi_\alpha } \right\rangle
\ \ \ \   (n \to \infty) .
\end{eqnarray}

	We work in configuration space, and adopt the usual
notation $\phi_\alpha^{(n)}({\vec q})\equiv \left\langle {\vec q} \right|
\left. {\phi _\alpha } \right\rangle^{(n)}$ where ${\vec q}$ signifies
the coordinate vector of the problem.
The straightforward evaluation of integrals in the
evolution equation is impractical except for the case of very low
dimension.  Instead, we consider, for given $n$ and $\alpha$, 
a set of vectors $\{{\vec q}_{\alpha i}^{\,(n)}\} (i=1 ... M)$ 
distributed according to the probability distribution
\begin{eqnarray}
\label{10}
P(\vec q)=\left| { \phi _\alpha ^{(n)}(\vec q) } \right|
  \cdot \left\{ {\ \int {d\vec q 
  \left| { \phi _\alpha ^{(n)}(\vec q) } \right| }} \right\}^{-1}
\end{eqnarray}
and try to represent the wave function by the method of
importance sampling.  
The sign information of $\phi _\alpha ^{(n)}(\vec q)$, 
which is missing in eq. (\ref{10}), can be recovered
by defining a real number
\begin{eqnarray}
\label{11}
s_{\alpha i}^{(n)} 
&\equiv& sgn\left({\phi_\alpha ^{(n)}(\vec q)} \right) \times
\\ \nonumber
& &
  \int {d\vec q \left| {\ \phi _\alpha ^{(n)}(\vec q) } \right|}
  \cdot 
\left\{ {\sum\limits_i {\left| 
                { \phi _\alpha ^{(n)}(\vec q_{\alpha i}^{\,(n)}) 
              } \right|}} \right\}^{-1}
\end{eqnarray}
for each point $\vec q_{\alpha i}^{\,(n)}$.  One can evaluate any
integral involving the wave function $\phi _\alpha ^{(n)}(\vec q)$
using the set $\{\vec q_{\alpha i}^{\,(n)},s_{\alpha i}^{(n)}\} \ 
(i = 1 ... M)$.  The most relevant example is the coordinate 
representation of the evolution equation eq. (\ref{8}), which now reads
\begin{eqnarray}
\label{12}
\phi_\alpha ^{(n+1)}(\vec q)
&=& \int { d\vec q\,' K_{\Delta \tau }(\vec q,\vec q\,')
                    \phi_\alpha ^{(n)}(\vec q) }
\\ \nonumber
& &
  - \sum\limits_{\beta =1}^{\alpha -1} 
       { \int { d\vec q\,' K_{\Delta \tau }(\vec q,\vec q\,')
                         \phi_\beta ^{(n)}(\vec q) 
       }      } \cdot 
  {{\lambda_{\beta \alpha }} \over {\Lambda _{\beta \beta }}} \\ \nonumber
&\approx& \sum\limits_i {K_{\Delta \tau }(\vec q,\vec q_{\alpha i}^{\,(n)})
                        s_{\alpha i}^{(n)}}
\\ \nonumber
& &
         -\sum\limits_{\beta =1}^{\alpha -1} 
             {\sum\limits_i 
                    {K_{\Delta \tau }(\vec q,\vec q_{\beta i}^{\,(n)})
                     s_{\beta i}^{(n)} 
                      }}\cdot 
          {{\lambda _{\beta \alpha }} \over {\Lambda _{\beta \beta }}} ,
\end{eqnarray}
where $K_{\Delta \tau }(\vec q,\vec q\,')\equiv \left\langle {\vec q} 
\right| \exp (-\Delta \tau  H) \left| {\vec q\,'} \right\rangle$
 is the step propagator in $q$-representation.  The overlap matrix 
$\lambda_{\beta\gamma}$ is given by
\begin{eqnarray}
\label{13}
\lambda _{\beta \gamma }
&\equiv& \ {}^{(n)}\!\left\langle {\phi _\beta } 
  \right| e^{-\Delta \tau H} \left| {\phi_\gamma } \right\rangle^{(n)}
\\ \nonumber
&\approx& \sum\limits_{i,j} 
  {K_{\Delta \tau }(\vec q_{\beta i}^{\,(n)},\vec q_{\gamma j}^{\,(n)}) 
  s_{\beta i}^{(n)}\ s_{\gamma j}^{(n)}} .
\end{eqnarray}
Another overlap matrix $\Lambda _{\beta \gamma } \equiv 
{}^{(n)}\!\left\langle {\phi_\beta } \right|\left. {\phi_\gamma } 
\right\rangle^{(n+1)}$
is calculated with the recursion relation
\begin{eqnarray}
\label{14}
\Lambda _{\beta \gamma }
=\lambda _{\beta \alpha}
+\sum\limits_{\alpha =1}^{\gamma -1} 
   {\Lambda _{\gamma \alpha }
    {1 \over {\Lambda_{\alpha \alpha }}}
   }\lambda _{\alpha \gamma } .
\end{eqnarray}
Since we always take care of the normalization of the wave functions
explicitly, the overall factor in the integrals of eq. (\ref{12}) is
irrelevant, and one can discard the unseemly constant (integral and
sum) in the definition eq. (\ref{11}) to regard $s_{\alpha i}^{(n)}$ as
simple sign, $s_{\alpha i}^{(n)}= \pm 1$.  The standard Metropolis
algorithm \cite{MR53} can be applied on 
eq. (\ref{12}) to obtain a new set of
numbers $\{q_{\alpha i}^{(n+1)},s_{\alpha i}^{(n+1)}\}$ 
which samples the $\phi _\alpha ^{(n+1)}(\vec q)$
from the set
$\{q_{\alpha i}^{(n)},s_{\alpha i}^{(n)}\}$ by a random walk.  
At each step, one can evaluate the energy integral 
with the sampling points and
associated sign as
\begin{eqnarray}
\label{15}
E_\alpha ^{(n)} 
&\equiv& {{\int {d\vec q \phi _\alpha ^{(n)}(\vec q) 
  H \phi _\alpha^{(n+1)}(\vec q)}} \over 
     {\int {d\vec q \phi_\alpha ^{(n)}(\vec q) 
      \phi_\alpha^{(n+1)}(\vec q)}}}
\\ \nonumber
&\approx& {{\sum\limits_{i,j} {HK_{\Delta \tau }
     (\vec q_{\alpha i}^{\,(n)},\vec q_{\alpha j}^{\,(n)})
         s_{\alpha i}^{(n)}s_{\alpha j}^{(n)}}} \over 
      {\sum\limits_{i,j} 
     {K_{\Delta \tau }(\vec q_{\alpha i}^{\,(n)},\vec q_{\alpha j}^{\,(n)}) 
        s_{\alpha i}^{(n)}s_{\alpha j}^{(n)}}}}
\end{eqnarray}
where $HK_{\Delta \tau }(\vec q,\vec q\,')\equiv \left\langle {\vec q} 
\right| H \exp (-\Delta \tau H) \left| {\vec q\,'} \right\rangle$.
This sequence of integrals, after the convergence, which we assume to 
occur before 
the $n_0$-th step, should yield the true eigenvalue
\begin{eqnarray}
\label{16}
E_\alpha  = \lim \limits_{N_s\to \infty } {1 \over N_s} 
\sum\limits_{n=n_0}^{n_0+N_s} {E_\alpha ^{(n)}} .
\end{eqnarray}
After the $n_0$-th step, all $N_s*M$ sets \{$\vec q_{\alpha i}^{\,(n)},
s_{\alpha i}^{(n)}$ \} 
($i = 1 ... M, n = n_0 ... n_0 + N_s$) can
be thought of as sampling the true eigenstate wave function that can
be used to calculate any observable.  It should be noted that the
eq. (\ref{12}) is exact only in the first line, 
and the second line has the
sampling error in the estimation of the overlaps, and includes the
contamination by the lower eigenstates.  However, because of the law
of large numbers, the fluctuating error at each step should cancel
each other in a long run, as long as the sequence of the state
$\phi_\alpha^{(n)}$ converges to a fixed limit.  
Therefore, in large $N_s$ limit, $E_\alpha$ of eq. (\ref{16})
does converge toward the true eigenvalue for any excited states,
albeit more slowly for higher states.

	Now we turn to the numerical examples.  Here, we can only
outline the results deferring the full detail to a forthcoming
publication.  First, we look at the one-dimensional motion of
a particle in a Morse potential to see the workings of the 
method in a simplest setting.  We
take the Hamiltonian \cite{MF53}
%
\begin{table}
\caption{
The energy eigenvalue of the first five states of the Morse 
system.  The raw [a] is the results with 
$\Delta\tau$ = 0.5, $M$ = 200 and $N$ = 400, 
while [b] is with $\Delta\tau$ = 0.2, $M$ = 
1000 and $N$ = 80. }
\begin{tabular}{|c|ccccc|} 
 $\alpha$&   1   &   2   &   3   &   4   &   5   \\ \tableline
  E [a]  & -7.00 & -5.29 & -3.90 & -2.86 & -1.84 \\ \tableline
  E [b]	 & -7.01 & -5.27 & -3.79 & -2.56 & -1.53 \\ \tableline
 \ \ (Exact)\ \  &(-7.031)&(-5.281)&(-3.781)&(-2.531)&(-1.531) \\ 
\end{tabular}
\end{table}
\begin{eqnarray}
\label{17}
H = -{1 \over 2}{{\partial ^2} \over {\partial q^2}}
    +8 e^{-q/2} -16 e^{-q} .
\end{eqnarray}
In Table 1, the first five eigenvalues evaluated with two sets of
sampling parameters ($M = 200, N_s = 400, \Delta\tau = 0.5$) and 
($M = 1000, N_s = 80, \Delta\tau = 0.2$) are tabulated.  
Statistical errors are estimated to be
0.2\% at most for both cases.  
One can see, in the table, that the
results do converge toward the analytical value 
$E_\alpha = -(17- 2\alpha)^2 / 32 $ with
larger $M$ and smaller $\Delta\tau$.  
%
\begin{figure}
\center\psbox[scale=0.42]{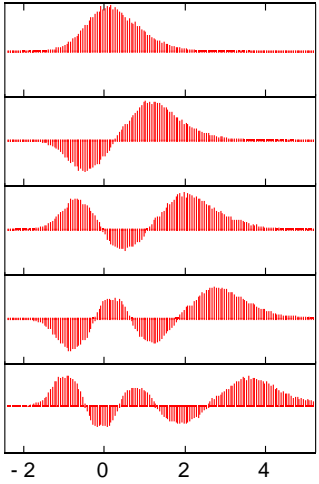}
\caption{
The profile of the first five wave function calculated with 
Monte Carlo technique.}
\end{figure}
In Fig. 1, we plot the wave functions,
which are obtained as suitably normalized probability 
distributions of sampling points taken separately for 
$s_{\alpha i}^{(n)}= 1$ and $s_{\alpha i}^{(n)} = -1$.
Their satisfactory quality is visually evident.

	We look at another numerical example of two identical
particles in a common two-dimensional asymmetric harmonic 
potential, mutually interacting through a Gaussian force.
This can be regarded as a simplified, but non-trivial prototype of
the deformed shell model \cite{NI55}.  The model Hamiltonian is
\begin{eqnarray}
\label{18}
H &=& \sum\limits_{k=1}^2 { {1 \over 2} }
  \left\{ {-{{\partial ^2} 
         \over {\partial x_k^2}}
  -{{\partial ^2} \over {\partial y_k^2}}
  +x_k^2+{{25} \over {16}}y_k^2} 
  \right\}
\\ \nonumber
& &
+v_0\exp 
  \left\{ {-{{(x_1-x_2)^2+(y_1-y_2)^2} \over {a^2}}} \right\} .
\end{eqnarray}
The coordinate vector 
$\vec q = (\vec q(\!1\!), \vec q(\!2\!)) 
= (x_1, y_1, x_2, y_2)$
is four dimensional.  We assume the particles to be fermions of
1/2 spin.  The spin degree of freedom and the identity of the
particle pose us the question of exchange symmetry.  We consider
a generic $S_z = 0$ state which we write as
\begin{eqnarray}
\label{19}
\phi_{ud}(\vec q(\!1\!),\vec q(\!2\!)) u_1 d_2 
+ \phi_{du}(\vec q(\!1\!),\vec q(\!2\!)) d_1 u_2
\end{eqnarray}
where $u$ and $d$ represent the $S_z = 1/2$ and $Sz = -1/2$ states of
individual particles respectively.  From eq. (\ref{19}), 
we can construct the totally antisymmetric state
\begin{eqnarray}
\label{20}
  \left\{ {\phi _{ud}(\vec q(\!1\!),\vec q(\!2\!))
          -\phi _{du}(\vec q(\!2\!),\vec q(\!1\!))} \right\} u_1 d_2
\\ \nonumber
+ \left\{ {\phi _{du}(\vec q(\!1\!),\vec q(\!2\!))
          -\phi _{ud}(\vec q(\!2\!),\vec q(\!1\!))} \right\} d_1 u_2 
\end{eqnarray}
which contains all the physical states.
Let us assume that the wave functions 
$\phi_{ud}$ and $\phi_{du}$ are sampled
respectively by the sets 
$\{ \vec q_{ud}(\!1\!)_i, \vec q_{ud}(\!2\!)_i, s_{udi} \}$ 
$(i = 1 ... M)$ and 
$\{ \vec q_{du}(\!1\!)_i, \vec q_{du}(\!2\!)_i, s_{dui} \}$ 
$(i = 1 ... M)$.  
Clearly, one can sample each term of eq. (20) by enlarging the
sets to include $2M$ elements with appropriate signs: 
Namely, for the first term, 
$\{\vec q_{ud}(\!1\!)_i, \vec q_{ud}(\!2\!)_i, s_{udi}\} \oplus 
 \{\vec q_{du}(\!2\!)_i, \vec q_{du}(\!1\!)_i,-s_{dui}\}$ 
$(i = 1 ... M)$, 
and for the second, 
$\{\vec q_{du}(\!1\!)_i, \vec q_{du}(\!2\!)_i, s_{dui}\} \oplus
 \{\vec q_{ud}(\!2\!)_i, \vec q_{ud}(\!1\!)_i,-s_{udi}\}$ 
$(i = 1 ... M)$.  
The operation of enlarging the sampling points for antisymmetrization 
should be carried out at each step of the evolution in
eq. (\ref{12}).  Otherwise, the system would pickup the small
contamination of other symmetries, and quickly evolve into the lowest
energy state of any symmetry, namely the totally symmetric state.
This scheme for the treatment of the exchange symmetry is
generalizable, in principle, to the system with 
arbitrary number of particles.  

The results of the calculation for 
the Hamiltonian eq. (\ref{18}) is
summarized in Table 2.  
The values for the interaction
parameters are taken to be $v_0 = 1$ and $a = 0.5$.
The sampling number $2M$ is set to 1600.  The
time step is chosen to be $\Delta\tau$ = 0.5.  The sampling point from
$N_s$ = 40 iterations (after the initial iterations of several
hundred) are included.  The first line is the energy eigenvalue of
first five states.  The second line lists the value of $4S(S+1)$ which
should serve as a barometer for the quality of wave functions
with respect to the separate spatial and spin symmetries: It should be
0 for spatially symmetric $S$ = 0 states, and 8 for spatially
antisymmetric $S$ = 1 states.
It appears that the third state still contains the contamination by
components of wrong symmetry by almost 10\%.
This  suggests that the choice
of either $\Delta\tau$ or $2M$ is still insufficient to obtain the
full convergence.  But the overall feature of the eigenstates are 
reasonably well captured even at this level.
The one-body density profiles calculated from the sampled points 
are depicted in Fig. 2.
Because of the repulsive two body force, 
the spatially symmetric states are pushed up in the energy 
compared to the antisymmetric companion states.
%
%
\begin{table}
\caption{
The energy eigenvalue and the total spin of 
the first five states of the Gaussian coupled two 
particles in two-dimensional oscillators.  The truncation parameters
are $\Delta\tau = 0.5$ and $2M = 1600$.}
\begin{tabular}{|c|ccccc|}
     $\alpha$   &    1  &   2   &   3   &   4   &   5   \\ \tableline
     E          &  2.44 &  3.34 &  3.43 &  3.56 &  3.69 \\ \tableline
\ \ $4S^2$ \ \  &  0.11 &  7.56 &  0.73 &  7.27 &  0.28 \\
\end{tabular}
\end{table}
%
\begin{figure}
\center\psbox[scale=0.42]{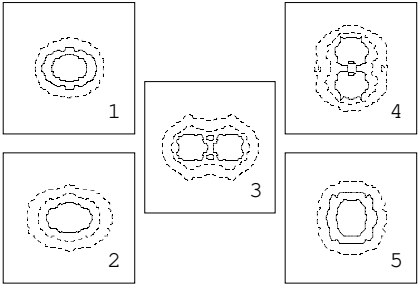}
\caption{
The one body density profiles of the first five eigenstates 
of the Hamiltonian eq. (18).}
\end{figure}

	Finally, we look at a linear array of oscillators whose
Hamiltonian is given by

\begin{eqnarray}
\label{21}
H=\sum\limits_{k=1}^N {\left\{ {{1 \over 2}\Pi_k^2+{1 \over {2a^2}}
(\phi_{k+1}-\phi_k)^2+U(\phi_k)} \right\}}a
\end{eqnarray}
where $\phi_k$ and $\Pi_k=\dot \phi_k$ are the amplitude and the
conjugate momentum of the $k$-th oscillator, and $a$ is 
the common distance between the neighboring oscillators.
The coupling between adjacent oscillator is 
given by a trigonometric form
\begin{eqnarray}
\label{22}
U(\phi )={m \over {\beta ^2}}\left( {1-\cos \beta \phi } \right) .
\end{eqnarray}
This is a discretized version of the one-dimensional sine-Gordon field.
The eigen spectra with the periodic boundary condition
\begin{eqnarray}
\label{23}
\phi _{N+1}=\phi _1
\end{eqnarray}
give the bosonic excitation modes.  There are also soliton modes which are 
identified as the fermionic excitations.  They are given by the twisted
boundary condition
\begin{eqnarray}
\label{24}
\phi _{N+1}=\phi_1+n_{\!f}{{2\pi}\over\beta}
\end{eqnarray}
where an integer $n_{\!f}$ is the number of kinks, or fermions.
Direct extraction of excitation spectra from the Hamiltonian,
eq. (\ref{21}) is difficult, because of the existence of 
translational zero mode of the system.
We define an operator $T$ which shifts the whole system by $a$ as
\begin{eqnarray}
\label{25}
& &T_a \Psi (\phi_1,\phi_2, ... , \phi_{N-1},\phi _N)
\\ \nonumber
&\equiv&  \Psi (\phi_2,\phi_3, ... , \phi_N,\phi _1+n_{\!f}{{2\pi}\over\beta})
\end{eqnarray}
where $\Psi$ is an arbitrary wave function.
Clearly, $T$ commutes with the Hamiltonian $H$.
Therefore, from an arbitrary eigenstates $\Psi_{n_{\!f}, \alpha}$ of $H$, 
one can project out the translationally invariant component as
\begin{eqnarray}
\label{26}
\tilde \Psi_{n_{\!f}, \alpha}
=\left\{ {1+T_a+T_a^2+...+T_a^{N-1}} \right\}\Psi_{n_{\!f}, \alpha} .
\end{eqnarray}
The projected eigenstates, which are the solutions of the
eigenvalue problem
\begin{eqnarray}
\label{27}
H \tilde\Psi_{n_{\!f}, \alpha} (\phi_1 ...\phi_N)
=E_{n_{\!f}, \alpha} \tilde\Psi_{n_{\!f}, \alpha} (\phi_1 ...\phi_N),
\end{eqnarray}
are calculated with the iterated Monte Carlo 
sampling as before.
The only complication is that we need to include 
all $N$-cyclic permutations
in the set of sampling points at each step of the iteration
to ensure the projection, eq. (\ref{26}): Namely, if 
$\Psi_{n_{\!f},\alpha}$ is sampled by a set of $M$ vectors 
$\{ \phi_{1 i}, \phi_{2 i} ... \phi_{N i}, s_i \}\ ( i = 1 ... M )$, 
$\tilde \Psi_{n_{\!f},\alpha}$ can be sampled by
$NM$ vectors 
$\{ \phi_{1 i}, \phi_{2 i}, ... , \phi_{N i}, s_i \} \oplus
 \{ \phi_{2 i}, \phi_{3 i}, ... , \phi_{1 i}+n_{\!f}{{2\pi}\over\beta},
     s_i \} \oplus ... \oplus
 \{ \phi_{N i}, \phi_{1 i}+n_{\!f}{{2\pi}\over\beta}, ... , 
    \phi_{N-1 i}+n_{\!f}{{2\pi}\over\beta},
     s_i\} \ ( i = 1 ... M )$.
From the eigenvalues {$E_{n_{\!f}, \alpha}$}, we can extract the mass
of bosons as
\begin{eqnarray}
\label{28}
m_{b1}=E_{0, 2}-E_{0, 1}, \ \ \ m_{b2}=E_{0, 3}-E_{0, 1}, \ \ \ etc..
\end{eqnarray}
where $m_{b1}$ is the mass of the elementary boson, $m_{b2}$, the
second boson, etc.. 
The mass of the fermion is given by
\begin{eqnarray}
\label{29}
M_{\!f}=E_{1, 0}-E_{0, 0} .
\end{eqnarray}
%
%
\begin{figure}
\center\psbox[scale=0.55]{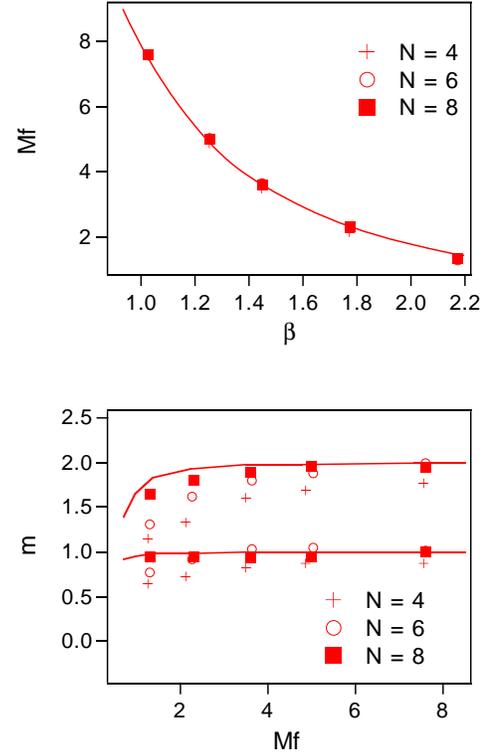}
\caption{
(a) The soliton mass $M_{\!f}$ as the function of the coupling 
$\beta$ and (b) the bosonic masses $m_{b1}$ and $m_{b2}$
as the function of soliton mass $M_{\!f}$ in
one dimensional sine-Gordon model.}
\end{figure}
In Fig.3 (a), we plot the fermion mass $M_{\!f}$ calculated with 
the Green function Monte Carlo procedure as the function of the
coupling constant $\beta$.  
In Fig. 3 (b), the boson masses $m_{b1}$ and $m_{b2}$ 
are plotted as the function of the fermion mass $M_{\!f}$.
The time step is taken to be $\Delta\tau = 0.5$. The 
sampling number $NM$ and the iteration number $N_s$ 
are chosen to be $360N$ and $400$
respectively.  Results for $N = 4, 6, 8$ are
shown in the figures. In Fig. 3 (a), results for different $N$
are practically indistinguishable. 
For the continuum limit $N \to \infty$, the mass spectra of the 
sine-Gordon model has been calculated within semi-classical 
approximation using the trace formula \cite{DH75}.  
This analytical result is shown in the figures as the line.  
The agreement with the numerical results seems to support the 
assertion by the authors of ref. \cite{DH75} that their 
results are exact in full quantum sense.  

	In above examples, the computational time required for the
convergent calculation still tends to exceed the time required for the
conventional direct diagonalization approach.  However,
the expected increase in time and storage for lager degrees
of freedom is more modest in the Monte Carlo calculation than 
in conventional approaches.
Also, the Monte Carlo methods have inherent affinity to 
the computational parallelism.  We therefore
believe that the results obtained here are sufficiently 
encouraging for the prospect of tackling more realistic 
problems both in many-body quantum theories and field theories.

	We express our gratitude to Drs. O. Morimatsu, M. Takizawa,
and T. Kawai for the helpful discussions.

%
\end{document}